# Traceable Dosimetry for MeV Ion Beams

G. Garty,[a,1] A.D. Harken[a] and D.J. Brenner[a]

*[a] Radiological Research Accelerator Facility, Columbia University*
*136 S. Broadway, Irvington, NY 10533, USA*

**ABSTRACT:**

Standard dosimetry protocols exist for highly penetrating photon and particle beams used in the clinic and in research. However, these protocols cannot be directly applied to shallow penetration MeV-range ion beams. The Radiological Research Accelerator Facility has been using such beams for almost 50 years to irradiate cell monolayers, using self-developed dosimetry, based on tissue equivalent ionization chambers.

To better align with the internationally accepted standards, we describe implementation of a commercial, NIST-traceable, air-filled ionization chamber for measurement of absorbed dose to water from low energy ions, using radiation quality correction factors calculated using TRS-398 recommendations. The reported dose does not depend on the ionization density in the range of 10-150 keV/μm.

## 1. Introduction

The Radiological Research Accelerator Facility (RARAF, www.raraf.org) at Columbia University is a radiation research facility that has been in operation since the late 60s. Our overall mission is to study the basic mechanisms of radiation effects in biological systems. We provide irradiations with neutrons, ions and electrons in the MeV range. Central to our work is robust, reproducible beam diagnostics, quantifying beam size and uniformity, LET, and delivered dose.

Over the past few years there has been much discussion of the inadequacy of dosimetry procedures used in radiation biology [1]. The turnkey nature of modern irradiators allows radiation biology studies to be designed and performed with minimal input from a radiation or medical physicist. This often results in inaccurate dosimetry as the users are typically unaware of important physical principles that modulate the dose delivered and, in some cases, rely on out-of-date dosimetry, performed when the irradiator was first installed, years before. This is evident from reviewing the literature where many papers are lacking important information needed to reproduce the described experiments [2].

While Radiation Oncology has long had standardized dosimetry protocols [3-5][2], ensuring that the dose delivered to the patient is indeed the intended one, most research laboratories do not follow these protocols, which tend to be elaborate and require specialized equipment, but result in dose uncertainties on the level of 1%, several times better than can be achieved using simpler methods.

These standards also pertain only to clinically relevant beams: The AAPM standards cover Orthovoltage X rays above 40 kVp [4], or energetic electrons and photons [3]. The IAEA standard [5] also covers energetic proton and ion beams, although it only addresses energies of clinical relevance, much higher than those available at RARAF.

We describe here modifications made to our routine beam diagnostics measurements [6], first developed in the 1970s, to conform to current recognized standards on dosimetry, and allowing NIST traceability and harmonization with other labs. In general, the standards needed to be extended to encompass the low energy ions available to us. This was done in the spirit of the original standard. As part of this process, we have transitioned from reporting kerma in tissue equivalent gas to the more prevailing standard of absorbed dose in water.

---

[1] Corresponding author. *E-mail*: gyg2101@cumc.columbia.edu

[2] The American Association of Physicists in Medicine (AAPM) is working on an equivalent standard for radiobiological research via Task Group 319.

**1.1 Overview of Existing Standards for Dosimetry**

Radiotherapy has long established standard protocols for absolute radiation dosimetry. In the US, these are the AAPM TG-51 [3] for high energy photon and electron beams and TG-61[4] for orthovoltage beams (40-300 kVp); there is no US standard for ion beam dosimetry. Internationally, IAEA Technical Report 398 [5] describes the code of practice for external beam dosimetry and also covers protons and heavy ions at clinically relevant energies.

Generically, these standards recommend utilization of ion chambers calibrated in absorbed dose to water at high energies and in air kerma at low energies. As the ion chamber is calibrated using a low LET photon beam (typically $^{60}$Co) a correction factor, due to beam quality needs to be applied. Measurements are generally performed within a water phantom using a broad beam.

The use of a gas filled ion chamber for measuring dose is predicated on the ability to translate energy deposited in the gas to the energy that would be deposited in bulk tissue. This can be done using Bragg-Gray cavity theory [7, 8]. Assuming that:

- the size of the cavity is small with respect to the range of the radiation traversing it, such that the cavity does not change the number, energy or direction of the particles traversing it, and
- the absorbed dose in the cavity is entirely due to charged particles traversing it,

then the dose in material, $D_M$, can be related to the dose deposited in the cavity $D_c$:

$$D_M = D_c \frac{(S/\rho)_M}{(S/\rho)_C}$$

where $S/\rho$ is the mass-electronic stopping power, averaged over all particles traversing the cavity.

TRS 398 [5] as well as the AAPM protocols [3, 4] recommend using cylindrical ion chambers within a water phantom for all measurements. However, they concede that this may not be appropriate for low energy beams (e.g. sub 100 kVp orthovoltage X rays), as the beam degradation, before reaching the detector, would be significant. In those cases, the recommendation is to use a thin parallel plate chamber with sufficient material behind it to provide the correct backscattering. In the case of MeV ions, where the residual range of the beam on exiting the accelerator is less than 1 mm in water, it is obvious that a water phantom (or even just the buildup cap of the ion chamber) would absorb all beam before it reaches the detector. The detector must therefore be operated with an exposed thin window facing the beam. The detector itself is usually thick enough to stop the beam entirely and therefore no backscatter material needs to be added.

Beyond irradiation geometry, the dosimetry codes of practice list a range of correction factors that are required for converting the measured ion chamber charge to dose.

**Calibration Factor** – This is the main parameter used to convert charge collected in the ion chamber to absorbed dose in water ($N_{D,w}$) or to air kerma ($N_K$). This calibration factor is obtained (and verified annually) by irradiating the ion chamber in a well characterized radiation field, (typically a low LET photon field such as $^{60}$Co) traceable to a national standard, at an Accredited Dosimetry Calibration Lab (ADCL).

**Air Density correction** – this is a correction due to local variations in temperature and pressure, with respect to the conditions at the ADCL, resulting in the air density in the detector being different from the one used during calibration.

$$k_{T,P} = \frac{P_{Ref}}{P} \times \frac{T[°k]}{T_{Ref}[°k]}$$

**Radiation Quality** – An important factor in calibration of ion chambers is the radiation quality. It is well known that the ion chamber readout depends on the type of radiation used. The standard ADCL calibration for clinically used ion chambers is done using a $^{60}$Co beam, though some ADCLs also offer various other standardized orthovoltage fields (e.g. Table II in [4]). To convert this calibration to other radiation fields, a calculated correction factor is applied.

As long as the Bragg-Gray cavity theory is valid, the beam quality correction factor, for measuring radiation quality "Q", using an ion chamber calibrated using for example $^{60}$Co is defined [5] as :

$$k_{Q,Co} = \frac{(S_{w,air})_Q (w_{air})_Q p_Q}{(S_{w,air})_{Co} (w_{air})_{Co} p_{Co}}$$

Where $S_{w,air}$ are the Spencer-attix water/air stopping power ratios (with a cut off of 10 keV), $w_{air}$ is the (differential) mean energy expended in air per ion pair formed and $p$ are the perturbations, which include all departures from ideal Bragg-Gray cavity conditions and depend on the ion chamber geometry.

$k_Q$ can be calculated based on stopping power data from the literature, as we do here, or calculated from a Monte Carlo simulation (e.g. [9]).

**Recombination corrections** - When radiation traverses an ion chamber a cloud of electron ion pairs is formed in the detector and collected onto the electrodes. If the voltage applied to the ion chamber is not sufficiently high, some of the electrons and ions may recombine before they are collected, resulting in a lower measured current.

This could be overcome by applying higher voltage to the ion chamber, but most chambers are limited to a few hundred volts before they are damaged by discharge. While in most cases ~300V are sufficient for efficient electron collection, at high dose rates, where there are many electron-ion pairs, the efficiency of the detector may be degraded.

This can be corrected by measuring the response function of the detector as a function of bias, *M(V)*, and fitting to $\frac{1}{M(V)} = \frac{1}{M_\infty} + \frac{b}{V}$ ;the correction factor then becomes $k_s^{ini} = \frac{M_\infty}{M(V)}$.

This approach only works at "reasonable" dose rates. The PTW Advanced Markus ion chamber, for example, is rated to 1.5 kGy/sec. In some FLASH irradiation scenarios, however, instantaneous dose rates can be significantly higher (several Gy per sub-µsec pulse), making ion chamber-based dosimetry impossible [10, 11]. In those cases, alternative techniques such as film dosimetry or air fluorescence can be used.

**1.2 Alternatives to absolute dose**

An alternative approach for beam quantification in low energy ion beams is to measure fluence, namely, the number of particles traversing the sample. Dose can then be calculated from the beam area and LET spectrum. Measuring fluence can be done using CR39 track etch film [12], though this only provides a retroactive quantification, or using an in-line ion chamber or other beam monitor. We have used this approach historically for our broad beam irradiation platform and continue to use it for microbeam irradiations. The main problem with this approach is that it is not traceable to a primary standard and thus prone to systematic errors. Nevertheless, in some applications it is the only feasible method for beam quantification. For example, in single-cell microbeam irradiations, dose is meaningless as the volume (or mass) irradiated by a single particle track is poorly defined. In such a case fluence becomes the only meaningful parameter.

**2. The RARAF Track Segment Platform**

The Track Segment "mono-LET" irradiation platform at RARAF [6, 13], has been in operation for more than 45 years following its construction in the mid-1970s. This platform was designed around the irradiation of thin biological samples using proton, deuteron and helium beams. Ongoing work aims to expand the palette of available ions to include lithium, boron and carbon at energies of up to 5.5 MeV/u.

The ion beam from the RARAF accelerator is energy selected, using a set of magnetic dipoles, defocused and exits the vacuum system through a 2.9 µm thick Havar window (Fig. 1a). The window includes a wedge-shaped collimator resulting in a roughly 40 mm x 6 mm beam. To mimic a broad beam for irradiation and dosimetry purposes, samples are placed on a large wheel (Fig. 1b) and scanned across the beam. Samples are typically placed in a 30 mm diameter holder with 6 µm mylar bottom. Assuming a thin target, this results in a "segment" of a particle track, having nearly constant LET, traversing the sample.

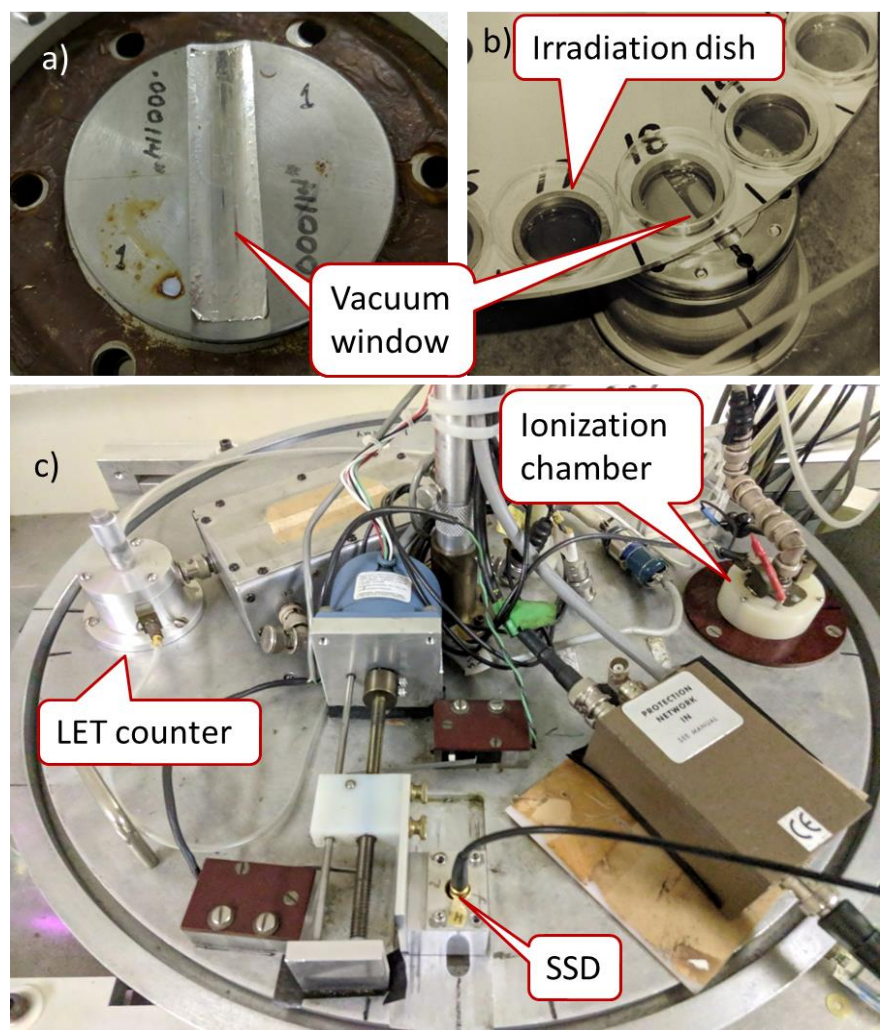

*Fig. 1: a) The havar exit window, b) irradiation wheel and c) dosimetry wheel at the Track Segment Platform with the 3 detectors mounted on the rotating wheel highlighted.*

**2.1 Dosimetry layout**

The dosimetry of the Track Segment platform is performed, prior to each experiment, using a combination of four detectors (Fig 1c):
1) A Lithium drifted Silicon solid state detector (**SSD**) for measuring the total energy of single particles and overall beam uniformity.
2) A tissue equivalent ionization chamber (**LET counter**) with an adjustable thickness (used to correct for day-to-day variations in temperature and pressure) measuring the LET spectrum of single particles.
3) A second, thin, **ionization chamber** for measuring of the total delivered dose in a simulated 6 µm thick tissue layer.
4) A **beam Monitor** placed inside the vacuum beam pipe. The monitor is a suppressed aperture that both shapes the beam and provides a real time current measurement of the wipe off, proportional to the amount of beam delivered to the sample (as long as the downstream beam focusing does not change).

In combination, these detectors measure all aspects of the chosen particles and the total energy delivered to a simulated 6 µm (cell monolayer) thickness. The first 3 detectors are all co-mounted on a rotational wheel,

Fig. 1, similar to the sample holder wheels, allowing fast switching between them for quick dosimetry measurements using the three in combination. This also guarantees that the detectors are all placed at the same distance from the beam exit window as the samples. The dosimetry and the irradiation protocols are carried out by a custom computer program that controls the wheel rotation stepper motor on an individual step basis determined by the dosimetry measurements and the beam current.

## 2.2 Beam Characterization at the Track Segment Facility

To characterize the beam, the three detectors are scanned across the havar window, under control of the beam monitor. This way all measurements are performed using the same "amount" of beam, regardless of any drift downstream of the monitor (e.g., in the ion source). This procedure is performed periodically during the experiment to check for potential slow drift in beam focusing (which we do not generally see).

### 2.2.1 Beam Uniformity Measurement

A Lithium drifted Silicon SSD (Ortec, Oak Ridge, TN), Fig. 2, with a ½ mm aperture, is used to record the energy spectrum of the ions that are coming out of the beam exit window. Additionally, the spatial uniformity of the beam is checked by scanning the silicon detector across the exit window at different positions along the length of the window. The total number of individual ions is counted by the detector during each scan. The variation in the number of ions registered at different positions reflects the variation in the beam intensity. This measurement allows adjusting magnet and focusing settings to obtain a uniform beam, centred on the sample to be irradiated.

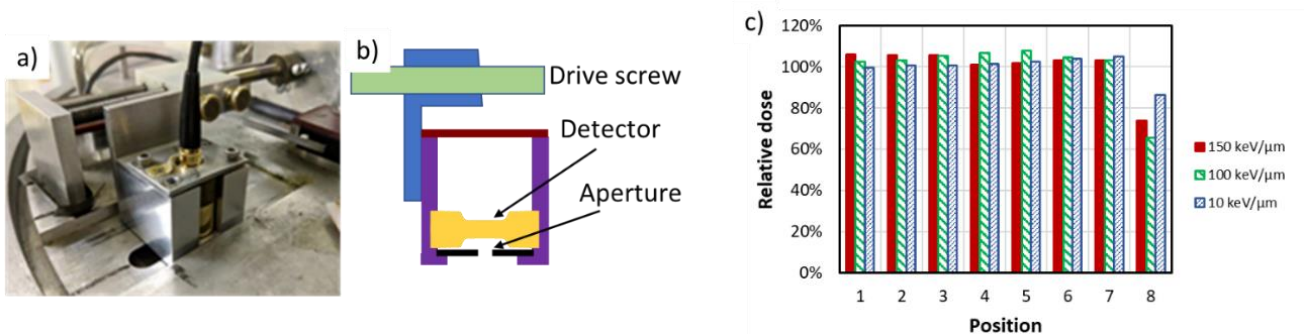

*Fig. 2: a) The Solid-State Detector (SSD) and b) a cross sectional view of its mount. The detector can be moved in and out on the wheel diameter to cover the whole beam window area for c) uniformity measurement, the distance between consecutive positions is 4.9 mm (3 mrad with respect to the 90º bending magnet).*

### 2.2.2 Beam Energy Measurement

This detector can also be used to verify beam energy. Fig 3 shows pulse height spectra from the surface barrier detector from two helium beams as well as from a spectroscopic, 100 nCi $^{148}$Gd alpha source (Eckert & Ziegler Isotope Products, Valencia, CA). $^{148}$Gd is a pure alpha emitter (3.2 MeV). As the source itself is not sealed, the resulting energy spectrum for alpha particles exiting the source should have a sharp edge at around 2.85 MeV (3.2 MeV, degraded by 3 mm of air) and a long tail on the low energy side corresponding to alpha particles self-attenuated by the gadolinium foil.

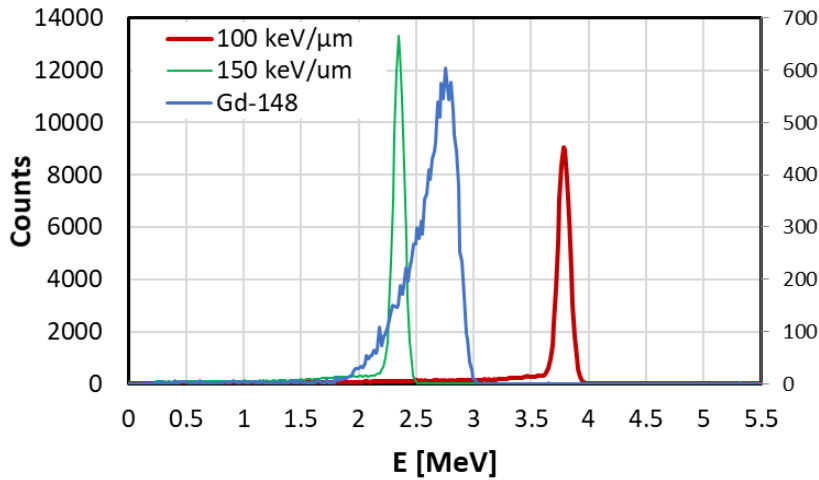

*Figure 3: pulse height spectrum from the surface barrier detector, for alpha particles having LET of 100 keV/μm (nominal energy of 4 MeV) 150 keV/μm (2.25 MeV) and from the gadolinium source (2.85 MeV).*

**2.2.3 LET measurement**

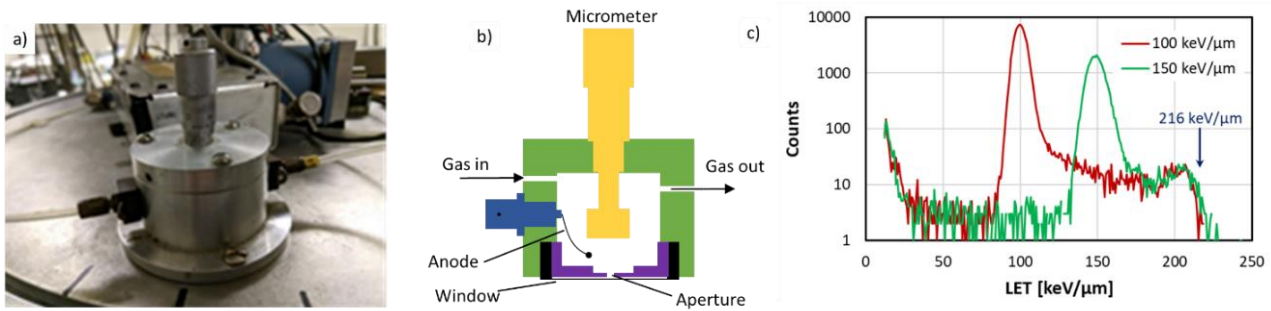

*Fig. 4: a) Photo and b) drawing of the Tissue Equivalent Ionization Chamber (TEIC). Note the stop driven by the micrometric screw to adjust the thickness of the ionization pathway. The unguarded readout is connected directly to a preamplifier. C) pulse height spectra from nominally 100 and 150 keV/μm helium ions.*

A unit gain tissue equivalent proportional counter (TEPC), Fig. 4, is used to measure the linear energy transfer (LET) of the ions. The counter is filled with methane-based tissue equivalent (TE) gas and operated in pulse mode, thus the energy deposited by each individual ion, while passing through the TEPC, is recorded as an LET spectrum (Fig 4c). The thickness of the TEPC collecting volume is manually adjusted by a micrometer spindle to represent the equivalent of a 6 μm thick muscle tissue, correcting for daily variations in temperature and atmospheric pressure. The upper limit of the measured spectrum corresponds to the maximum energy that the particle can deposit along a 6 μm path length. For example, a helium nucleus traversing 6 μm of tissue can deposit at most 1.3 MeV, thus the edge on the LET histogram would be at 216 keV/μm (Fig 5), with a different edge value for each ion type. Thus, the LET chamber is essentially self-calibrating and can give absorbed dose by providing both LET and fluence, independent of the more conventional instrumentation. As an example, Fig. 4c shows two LET spectra generated by 100 and 150 keV/μm He$^{++}$ beams.

A cross check of the LET is provided by dividing the dose measurement from the ionization counter by the number of particles counted by the LET/SSD counters, correcting for the aperture size on the different counters. This usually provides a consistent LET value. Discrepancy of more than 10% indicates problems with the ionization chamber.

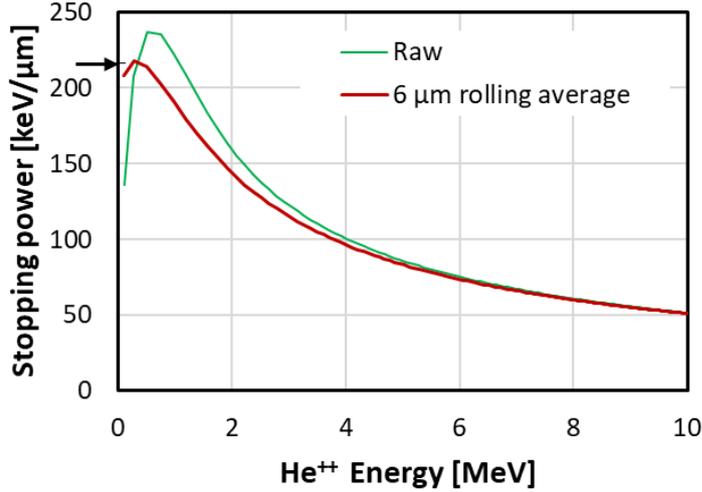

*Figure 5: stopping power for Helium ions in methane TE gas ([14], as tabulated in the NIST ASTAR database; Green) and averaged over the thickness of a 6 μm-equivalent ion chamber (red). 216 keV/μm is indicated by arrow.*

**2.2.4 Dose Measurement**

An ionization chamber, Fig. 6, also filled with methane-based TE gas, is used to measure dose. The guarded parallel plate ionization chamber is designed to mimic a 6 µm tissue layer.

Dose to tissue is calculated from the current collected by the ion chamber and read via a Kiethley 619 electrometer, using a calibration factor calculated from the detector volume, collection efficiency and *w* value for methane TE gas (30.5eV/ion pair):

$$N_{k,TE} = \frac{T[°k] \times 30.5}{193500 \times P[Torr]} \frac{Gy}{nC} .$$

Prior to each experiment, the ionization chamber is used to calibrate the beam current monitor which consists of a set of apertures that constantly wipe off a portion of the beam penumbra. The beam current monitor controls the rate at which the wheel with samples moves over the beam during irradiation. During the dosimetry protocol, the ionization chamber is moved over the beam exit aperture in the same way and its current output is integrated to give a total charge produced inside the collecting volume. This charge is a measure of the dose deposited to the tissue equivalent gas inside the chamber. After the initial scan, a new number of counts per step is calculated based on the correction factor which is equal to a ratio between the desired dose and the dose that was measured by the ionization chamber. This procedure is repeated until the correction factor converges to 1± 0.1%. At that point the dosimetry protocol is finished, and beam monitor counts are calibrated in terms of the absorbed dose. This calibration is then used to coordinate the exposure time of the samples, passage time over the beam exposure window, to the charged particle beam rate.

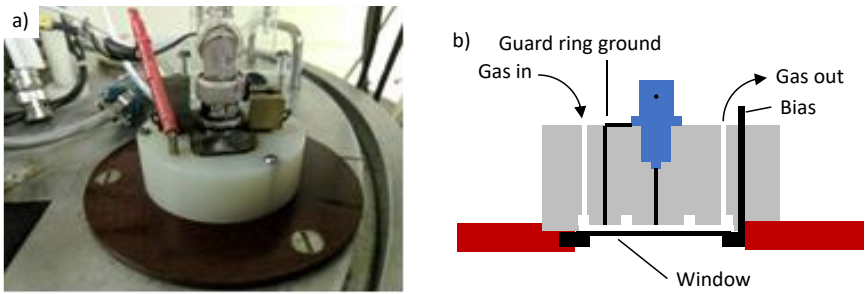

*Fig. 6: a) Photo and b) drawing of the Ionization Chamber.*

# 3. Calibrating Dosimetry to Traceable Standards

Current standards of radiation dosimetry require periodic calibration of the ion camber to a national standard and reporting absorbed dose to water when possible. To align the track segment dosimetry with these standards, we have integrated a commercial air filled Advanced Markus Ion Chamber (PTW, Germany)[15] into the dosimetry protocol. This allows annual removal of the ion chamber and calibration at an ADCL. To accommodate the short range of our ion beams, the 30 µm thick polyethylene entrance window was replaced with a 6 µm thick aluminized mylar window. The detector was then calibrated to absorbed dose in water using a Co-60 beam, by an ADCL (the M. D. Anderson Dosimetry Laboratory, Houston, TX) giving $N_{D,w} = 1.444 \; Gy/nC \pm 1.3\%$.

The chamber was operated at a voltage of -300V – we have seen that the reading at this voltage (at about 5 Gy/min of helium ions) is independent of voltage, indicating negligible recombination corrections. Correction for air humidity, typically on the order of 0.3% [16] was also not made.

Perturbation factors, correcting for departures from ideal Bragg-Gray cavity conditions are not, generally available for ion beams, however it has been shown that for electron beams, the Advanced Markus ion chamber has perturbation factors close to unity [17].

## 3.1 Calculation of $k_Q$

Table 1 shows the calculation of $k_Q$, based on values from the literature

For $^{60}$Co gamma rays $w_{air}$ is 33.97 eV ± 0.2% [5] and $S_{w,air}$ is 1.133 ± 0.2% [5]. For ion beams, things are a bit more complex:

In the general case for ion SOBP conditions [5] $S_{w,air}$ and $w_{air}$ need to be averaged over all particles and energies present in the beam, which is extremely difficult to achieve. In our scenario, where the beam is nearly monoenergetic and does not contain contaminating secondary fragments, stopping powers are easier to calculate by Monte Carlo. These values have been calculated for clinical proton [18] and carbon beams [19] and for other high energy beams [9] but data is sparse for lower energy monoenergetic beams, usually available as a by-product of high energy beam calculations [9]. It would be reasonable to assume that at low energies, the contribution of energetic delta rays to the stopping power would be small and so, in this work we will use the unrestricted stopping power, which is more readily available.

Based on data from ICRU report 90 [16], the calculated $S_{w,air}$ values for protons and helium ions range from 1.136 to 1.147 (about 1% lower than those we have previously calculated using ICRU 49 data[14]). The values for heavier ion beams, can be calculated based on values from ICRU report 73 [20](Li, B) and 90 [16] (C), resulting in $S_{w,air}$ values slightly higher than the recommended value of 1.13 for high energy (clinical) heavy ion beams [5]. Our heavy ion values are also slightly higher than those calculated using FLUKA by Luoni et al [9]. However, those simulations were optimized and validated for significantly higher energies. A 3-4% discrepancy, especially for low energy heavy ions (with a range in water comparable to the 0.2cm$^3$ scoring voxel size used), is not surprising.

For the relevant energies (>10 keV for protons) $w_{air}$ approaches the value of $W_{air}$ and is generally accepted as 34.40±0.14 eV for protons and 34.71±0.52 eV for carbon [16, 21]. Gray derived a formula for $w_{air}$ for alpha particles of energy E which fits experimental values well $w_{air} = 32.75 + \frac{6.5}{\sqrt{E[MeV]}}$ [22], although ICRU recommend using a slightly lower value of 35.08±0.7% for 5 MeV alpha particles [23]. If there are no values for heavier ions, it is recommended to use the ones for alphas at the same velocity [24].

Based on these values, and assuming unity perturbation factors, we have calculated $k_Q$ for the hydrogen and helium beams in current use and the heavier beams under development.

*Table 1 calculation of kQ factors for beams used in this work and for heavy ion beams under testing. Superscripts indicate source of values from literature: [*][14], [†][16], [&][20], [‡][22], [§]scaled from protons [16], [°]Scaled from Helium [22]; FLUKA-simulated $S_{w,air}$ values [•][9] provided for reference).*

| Ion | E [MeV] | LET [keV/µm] | W [eV/e-] | S/ρ Water [MeV cm²/g] | S/ρ Air [MeV cm²/g] | S/ρ TE [MeV cm²/g] | $S_{w,air}$ | kQ |
|---|---|---|---|---|---|---|---|---|
| $^{60}$Co | | | 33.97[†] | | | | 1.133[†] | 1 |
| H[+] | 4.5 | 9.4 | 34.4[†] | 85.86[†] | 75.53[†] | 88.8[*] | 1.136 / 1.14[•] | 1.016 |
| D[+] | 2.5 | 32.9 | 34.4[§] | 191.2[§] | 222.9[§] | 233.4[§] | 1.165 | 1.042 |
| He[++] | 10 | 70 | 34.8[‡] | 529.4[†] | 463.7[†] | 545.9[*] | 1.142 / 1.12[•] | 1.033 |
| He[++] | 8 | 100 | 35.1[‡] | 624.3[†] | 545.6[†] | 645[*] | 1.144 / 1.11[•] | 1.042 |
| He[++] | 7 | 150 | 35.2[‡] | 696.8[†] | 607.8[†] | 711.8[*] | 1.147 / 1.11[•] | 1.049 |
| | | | | | | | | |
| Li[3+] | 15 | 330 | 34.8[°] | 1247[&] | 1072[&] | 298.3[&] | 1.149 | 1.052 |
| B[5+] | 25 | 600 | 34.8[°] | 3322[&] | 2872[&] | 545.1[&] | 1.146 | 1.046 |
| C[6+] | 30 | 750 | 34.7[†] | 4636[&] / 4300[†] | 4019[&] / 3681[†] | 676.5[&] | 1.154 / 1.168 / 1.125[•] | 1.040 / 1.053 |

## 4. Measuring dose

We have irradiated our modified, Advanced Markus ion chamber on the track segment facility using five different beams, spanning the available LET range, at nominal doses of 0, 1, 2, 4, 6, 8 and 10 Gy and a dose rate of roughly 5 Gy/min. For each measurement, we also performed a sham irradiation, where the chamber was not above the beam, but the same nominal dose was delivered. This "background" value was typically 0.1% of the measured dose. The reported doses were corrected using the appropriate $k_Q$ value from table 1, as well as temperature and pressure correction (T=22.7°C; P=764.4 Torr; $k_{TP}$=0.997). Perturbation and recombination corrections were not made. Table 2 shows the uncertainty budget, based on reported uncertainties in w and S/r values[16] leading on an overall error of under 4% for protons and helium and 5.4% for carbon.

*Table 2: Error budget for ion dose estimation. Only type B (non-stochastic) errors are shown.*

| Source of error | Value | | | |
|---|---|---|---|---|
| | $^{60}$Co | protons | helium | carbon |
| AMIC calibration | 1.3% | | | |
| w | 0.2% | 0.4% | 2% | 1.5% |
| $S_{w,air}$ | 0.2% | 3% | 3% | 5% |
| Total: | | 3.3% | 3.8% | 5.4% |

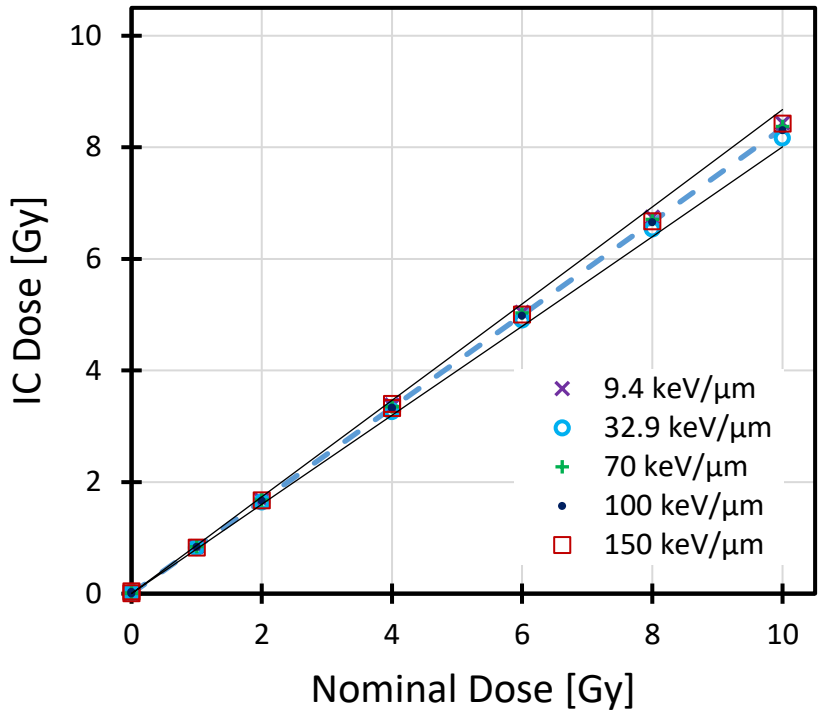

*Figure 7: Measured dose from the Advanced Markus ion chamber vs nominal dose reported by the TEPC. The dashed line represents the average of all LET values. The thin lines denote a ±4% error (Table 2).*

As expected, the measured dose response (Fig. 7) was highly linear ($R^2>0.999$), and importantly, independent of LET in the range tested.

The absorbed dose in water values are somewhat lower than the nominal dose values. This is partially because the TEPC was originally calibrated to dose deposited in tissue (or more correctly, methane-based TE gas), which has stopping power a few percent higher than that of water (see table 1) and partially due to potential drifts in the detector and electronics response, since they were first calibrated. Going forward RARAF is transitioning from reporting "dose to tissue" to reporting "absorbed dose in water", as per current AAPM and international standards.

## 5. Conclusions

Reliable, rigorous and reproducible dosimetry is a crucial aspect of radiobiology. We have developed protocols for NIST-traceable low energy ion dosimetry and demonstrated dosimetry using the RARAF charged particle track segment facility. Going forward we will used absorbed dose to water, as measured by a calibrated, NIST-traceable ionization chamber as our ground truth for dosimetry of the track segment facility, instead of kerma in TE gas, as used to date.

## 6. Acknowledgement

This work was funded by NCI grant # U01CA236554 to the Radiological Research Accelerator Facility. The content is solely the responsibility of the authors and does not necessarily represent the official views of the National Cancer Institute or National Institutes of Health.

## 7. References

1. Desrosiers, M., et al., *The Importance of Dosimetry Standardization in Radiobiology.* Journal of Research of the National Institute of Standards and Technology, 2013. **118**: p. 403.

2.  Draeger, E., et al., *A Dose of Reality: How 20 Years of Incomplete Physics and Dosimetry Reporting in Radiobiology Studies May Have Contributed to the Reproducibility Crisis.* International Journal of Radiation Oncology*Biology*Physics, 2020. **106**(2): p. 243-252.

3.  Almond, P.R., et al., *AAPM's TG-51 protocol for clinical reference dosimetry of high-energy photon and electron beams.* Medical Physics, 1999. **26**(9): p. 1847.

4.  Ma, C.-M., et al., *AAPM protocol for 40–300 kV x-ray beam dosimetry in radiotherapy and radiobiology.* Medical Physics, 2001. **28**(6): p. 868-893.

5.  International Atomic Energy Agency, *TRS 398 Absorbed Dose Determination in External Beam Radiotherapy: An International Code of Practice for Dosimetry Based on Standards of Absorbed Dose to Water*. 2000: Vienna.

6.  Bird, R.P., et al., *Inactivation of Synchronized Chinese Hamster V79 Cells with Charged-Particle Track Segments.* Radiation Research, 1980. **82**(2): p. 277-289.

7.  Bragg, W.H., *Studies in Radioactivity*. 1912, London: MacMillan and Co. ltd.

8.  Gray, L.H., *An Ionization Method for the Absolute Measurement of γ -Ray Energy.* Proceedings of the Royal Society of London. Series A, Mathematical and Physical Sciences, 1936. **156**(889): p. 578-596.

9.  Luoni, F., et al., *Beam Monitor Calibration for Radiobiological Experiments With Scanned High Energy Heavy Ion Beams at FAIR.* Frontiers in Physics, 2020. **8**.

10. Rahman, M., et al., *Electron FLASH Delivery at Treatment Room Isocenter for Efficient Reversible Conversion of a Clinical LINAC.* International Journal of Radiation Oncology*Biology*Physics, 2021. **110**(3): p. 872-882.

11. Garty, G., et al., *FLASH irradiator at the Radiological Research Accelerator Facility.* Scientific Reports 2022. **in Review**: p. Preprint available at: https://doi.org/10.21203/rs.3.rs-1281287/v1.

12. Chan, K.F., et al., *Alpha-particle radiobiological experiments using thin CR-39 detectors.* Radiation Protection Dosimetry, 2006. **122**(1-4): p. 160-162.

13. Marino, S.A., *50 Years of the Radiological Research Accelerator Facility (RARAF).* Radiation Research, 2017. **187**(4): p. 413-423.

14. Berger, M.J., et al., *ICRU Report 49: Stopping Powers and Ranges for Protons and Alpha particles.* Journal of the International Commission on Radiation Units and Measurements, 2016. **os25**(2): p. NP-NP.

15. Pearce, J., R. Thomas, and A. Dusautoy, *The characterization of the Advanced Markus ionization chamber for use in reference electron dosimetry in the UK.* Physics in medicine and biology, 2006. **51 3**: p. 473-83.

16. Seltzer, S.M., et al., *ICRU Report 90: Key data for ionizing Radiation Dosimetry: Measurment Standards and Applications.* Journal of the International Commission on Radiation Units and Measurements, 2016. **14**(1): p. NP-NP.

17. Voigts-Rhetz, P.v., H. Vorwerk, and K. Zink, *On the Perturbation Correction Factor pcav of the Markus Parallel-Plate Ion Chamber in Clinical Electron Beams* International Journal of Medical Physics, Clinical Engineering and Radiation Oncology 2017. **6**(2): p. 150-161.

18. Medin, J. and P. Andreo, *Monte Carlo calculated stopping-power ratios, water/air, for clinical proton dosimetry (50-250 MeV).* Phys Med Biol, 1997. **42**(1): p. 89-105.

19. Geithner, O., et al., *Calculation of stopping power ratios for carbon ion dosimetry.* Physics in Medicine and Biology, 2006. **51**(9): p. 2279-2292.

20. Bimboot, R., et al., *ICRU Report 73: Stopping of Ions Heavier than Helium.* Journal of the International Commission on Radiation Units and Measurements, 2005. **5**(1): p. 0-0.

21. International Atomic Energy Agency, *IAEA TecDoc 799: Atomic and Molecular Data for Radiotherapy and Radiation Research*. 1995, Vienna: INTERNATIONAL ATOMIC ENERGY

AGENCY.

22. Cranshaw, T.E. and J.A. Harvey, *MEASUREMENT OF THE ENERGIES OF α-PARTICLES.* Canadian Journal of Research, 1948. **26a**(4): p. 243-254.

23. International commission on radiation units and measurements, *ICRU Report 31: Average Energy Required to Produce An Ion Pair.* Journal of the International Commission on Radiation Units and Measurements, 1979. **os-16**(2).

24. Binks, W., *Energy per Ion Pair.* Acta Radiologica, 1954. **41**(sup117): p. 85-104.